\documentclass[9pt,twocolumn,twoside]{opticajnl}

\journal{josab} 

\setboolean{shortarticle}{false}

\usepackage{amsmath,commath,siunitx,physics}
\newcommand{\SIadj}[2]{\SI[number-unit-product={\text{-}}]{#1}{#2}}

\newcommand{\ie}{\textit{i}.\textit{e}.,\ }
\newcommand{\eg}{\textit{e}.\textit{g}.,\ }
\newcommand*{\Exp}{\mathrm{e}}
\newcommand*\diff{\mathop{}\!\mathrm{d}}

\title{A simple accurate way to model noise-seeded ultrafast nonlinear processes}

\author[*,1]{Yi-Hao Chen}
\author[1]{Frank Wise}

\affil[1]{School of Applied and Engineering Physics, Cornell University, Ithaca NY 14853, USA}

\affil[*]{Corresponding author: yc2368@cornell.edu}

\begin{abstract}
Noise can play an important role in nonlinear pulse propagation. It is not only the origin of fluctuations in supercontinuum but can also determine the generated signal amplitude and phase, as seen in phenomena such as noise-seeded four-wave mixing (FWM) and spontaneous Raman scattering. Current models rely on input-pulse shot noise and Raman Langevin term to simulate noise-seeded Kerr effects and spontaneous Raman scattering, respectively. However, they cannot accurately simulate any nonlinear effects exhibiting asymmetries, for example, between spontaneous Stokes and anti-Stokes generation, or in amplification with emission and absorption cross sections. Moreover, combining these two techniques does not lead to accurate modeling of both effects simultaneously, especially in unsaturated Raman operational regimes or supercontinuum where spontaneous Raman emission noise plays a crucial role. To resolve this issue, we propose a new modified shot-noise model by directly introducing noise into the pulse propagation equation. It minimally modifies current models that contain only stimulated terms.
\end{abstract}

\setboolean{displaycopyright}{true}

\begin{document}

\maketitle

In general, nonlinear pulse propagation can be categorized into stimulated and spontaneous processes. Stimulated processes modulate existing fields through either intrapulse (\eg self-phase modulation \cite{Liu2016a,Chung2018} or soliton self-frequency shift \cite{Gordon1986,Mitschke1986}) or interpulse (\eg stimulated Raman scattering) nonlinear evolutions. On the other hand, spontaneous processes include nonlinear generation initiated from noise, such as spontaneous Raman scattering \cite{Mostowski1981,Raymer1981} or noise-seeded FWM \cite{Nagashima2019}. The spontaneously generated field later acts as a seed for the subsequent stimulated processes, potentially resulting in noisy stimulated nonlinear evolutions. This has been widely investigated in the context of, for example, supercontinuum, whose intense spectral broadening typically involves fields initiated from noise through noise-seeded Kerr effects or spontaneous Raman scattering \cite{Rampur2021}. To study these noise-involved processes accurately, it is crucial to take into account noise effects in numerical simulations.

Nowadays, the most widely-used model to simulate the noise is the inclusion of input-pulse shot noise \cite{Eggleston1980,Dudley2006,Frosz2010,Genier2019}, where the noise with an one-noise-photon spectral distribution $P_{\text{noise}}(\nu)=h\nu$ [with the unit of \si{\watt/\Hz}] is added to the input field. The propagation is governed only by stimulated terms such that noise-involved processes result from the stimulated evolution of the shot noise. This leads to the added spectral field $A_{\text{noise}}(\omega)=\abs{A_{\text{noise}}(\omega)}\mathfrak{R}(\omega)$, whose amplitude $\abs{A_{\text{noise}}(\omega)}=C_{\mathfrak{F}}\sqrt{T^wh\nu}$ ($T^w$ is the time window; see \ref{sec:Fourier_Transform}) and $\mathfrak{R}(\omega)$ is a random complex number, following the joint normal distribution over different frequencies. Fourier transform follows $\mathfrak{F}[A](\omega)=C_{\mathfrak{F}}\int_{-\infty}^{\infty}A(t)\Exp^{i\omega t}\diff t$. Similar to $C_{\mathfrak{F}}$, $C_{\mathfrak{IF}}$ is the constant in inverse Fourier transform with $C_{\mathfrak{F}}C_{\mathfrak{IF}}=1/\left(2\pi\right)$. Its discrete-Fourier-Transform (DFT) counterpart (if we also ignore the $\diff t$ factor when computing the spectral component) $\abs{A_{D,\text{noise}}(\omega)}=\frac{C_{\mathfrak{F}_D}}{\triangle t}\sqrt{T^wh\nu}$. DFT follows $\mathfrak{F}_D[A](\omega)=C_{\mathfrak{F}_D}\sum_{n=1}^{\mathfrak{N}}A(t_n)\Exp^{i\omega t_n}$, where $\triangle t$ and $\triangle\nu$ are temporal and spectral sampling periods, respectively. $C_{\mathfrak{F}_D}C_{\mathfrak{IF}_D}=1/\mathfrak{N}$ ($\mathfrak{N}$ is the number of sampling points). The Fourier-Transform convention follows the wave-vector physical formulation $A(t)\sim\int A(\omega)\Exp^{i(\beta z-\omega t)}\diff\omega$, leading to $C_{\mathfrak{F}_D}=\frac{1}{\mathfrak{N}}$ for the DFT. With this convention, $\abs{A_{D,\text{noise}}(\omega)}=\sqrt{h\nu\triangle\nu}$. Because of the random $\mathfrak{R}(\omega)$, different shot noises are introduced to the input field on a pulse-by-pulse basis, which accounts for its term ``shot noise.'' It is worth noting the shot noise discussed in this article focuses only on those that mimic vacuum fluctuations \cite{Milonni1994}, rather than ``real'' shot noise existing in the source, such as those produced in mode-locked lasers and amplifiers \cite{Frosz2010}.

Shot noise can model various physical phenomena, covering not only multiple $\chi^{(3)}$ nonlinear processes, from noise-seeded Kerr effects to spontaneous Raman scattering \cite{Chen2024}, but also amplification in a rare-earth-doped fiber amplifier. Specifically, amplified spontaneous emission (SE) in a rare-earth-doped fiber amplifier is modeled as the growth of one-noise-photon vacuum fluctuations \cite{Giles1991,Chen2023a,Chen2025}, which, in principle, can also be simply modeled as the amplification of the injected shot-noise field. With the one-photon-per-mode shot noise, all the noise-seeded spontaneous processes can be treated as stimulated emission/amplification of the shot noise, which significantly simplifies numerical simulations.

Despite the general applicability of shot noise across a broad range of noise-involved physical phenomena, in fact, it cannot accurately model some of the aforementioned processes. Spontaneous Raman generation exhibits spectral asymmetry due to the Stokes preference and thermal phonons $\left[n_{\text{th}}(\abs{\Omega})+\Theta(-\Omega)\right]$, where thermal phonons follow the Bose-Einstein distribution $n_{\text{th}}(\abs{\Omega})=\left[\Exp^{\hbar\abs{\Omega}/k_B\mathbf{T}}-1\right]^{-1}$ and $\Theta(-\Omega)$ is the Heaviside step function \cite{Chen2024}. However, stimulated Raman scattering of the shot noise does not display this effect. Additionally, stimulated Raman scattering induces Raman loss for the shot noise at anti-Stokes frequencies. Due to the incoherent nature of the spontaneous generation, phase-matching effects, such as FWM, are expected to have minimal influence. Noise fluctuates with random phases during optical propagation such that the phase-matching relation cannot be consistently satisfied. However, as shot noise, superimposed on the input field, remains invariant over a single propagation, it acquires a consistent phase increment governed by the optical environment. Consequently, its evolution can be significantly influenced by (stimulated) phase-matching processes, which cannot be compromised with pulse-to-pulse randomness, equivalently by averaging over multiple simulations. Phase match of vacuum fluctuations arises as a numerical artifact inherent to the shot-noise model. Amplification in a fiber amplifier involves both emission and absorption processes in rare-earth ions. However, SE arises solely from the emission processes. The absence of the absorption effect in an actual SE renders the application of shot noise inappropriate in this scenario, producing underestimated SE in the shot-noise approach due to its stimulated absorption effect. In general, we expect the spontaneous generation of new spectral components to grow from zero values. Addition of shot noise creates a numerically abrupt increase of the spectral intensity. Furthermore, if the light propagates over multiple fiber segments, there are ambiguous and potentially-multiple applications of the shot-noise technique, artificially elevating the noise.

To resolved the issue with input-pulse shot noise, noise contribution should instead be included directly in the governing evolution equation. This has been well studied in spontaneous Raman scattering, whose noise inclusion is achieved with the quantum-statistical Langevin term \cite{Kaertner1994,Drummond1996,Drummond2001,Dudley2006}. Furthermore, it is found that the computation of the Raman Langevin term is equivalent to the stimulated Raman scattering of the vacuum fluctuations with an inclusion of the Stokes preference and thermal phonons \cite{Chen2024}. Here, we extend this approach to the amplification and the Kerr nonlinearity, where SE and noise-seeded Kerr processes can both be treated as stimulated modulation of the vacuum fluctuations during pulse propagation. This leads to the following pulse propagation equation:
\begin{equation}
\partial_zA_p=\hat{\mathcal{D}}A_p+\left(\hat{\mathcal{G}}'+\hat{\mathcal{N}}^K+\hat{\mathcal{N}}^R\right)\left(A_p+A_{p,\text{noise}}(\omega)\right),
\label{eq:partial_Ap}
\end{equation}
where $A_{p,\text{noise}}(\omega)$ is the noise field with the one-noise-photon spectral distribution and a random spectral phase, pre-determined before the propagation starts. $\hat{\mathcal{D}}$, $\hat{\mathcal{G}}'$, $\hat{\mathcal{N}}^K$, and $\hat{\mathcal{N}}^R$ represent the dispersion, gain, Kerr and Raman nonlinear operators, respectively. Eq.~(\ref{eq:partial_Ap}) can be considered a \emph{modified shot-noise approach}, where shot noise is considered only in the governing equation. However, $\hat{\mathcal{G}}'$ differs from its conventional stimulated $\hat{\mathcal{G}}$ due to additional consideration of the SEs. In the general multimode formulation with unidirectional pulse propagation equation (UPPE) [Eq.~(\ref{eq:UPPE})] \cite{Kolesik2004,Poletti2008,Travers2011}, these operators follow
\begingroup \allowdisplaybreaks 
\begin{subequations}
\begin{align}
\hat{\mathcal{G}}'\left(A_p+A_{p,\text{noise}}\right) & =\hat{\mathcal{G}}A_p+\hat{\mathcal{G}}_eA_{p,\text{noise}} \allowdisplaybreaks \\
\hat{\mathcal{N}}^K\left(A_p+A_{p,\text{noise}}\right) & =\hat{\mathcal{C}}^K_{p\ell mn}\left(A_{\ell}A_mA_n^{\ast}\right) \label{eq:NK} \nonumber \\
&\hspace{-9.6em} +\hat{\mathcal{C}}^K_{p\ell mn}\left(A_{\ell,\text{noise}}A_mA_n^{\ast}+A_{\ell}A_{m,\text{noise}}A_n^{\ast}+A_{\ell}A_mA_{n,\text{noise}}^{\ast}\right) \allowdisplaybreaks \\
\hat{\mathcal{N}}^R\left(A_p+A_{p,\text{noise}}\right) & =\hat{\mathcal{C}}^R_{p\ell mn}\left(A_{\ell},A_mA_n^{\ast}\right) \nonumber \\
&\hspace{-2em} +\hat{\mathcal{C}}^R_{p\ell mn}\left(A_{\ell,\text{noise}},A_mA_n^{\ast}\right) \nonumber \\
&\hspace{-2em} +\hat{\mathcal{C}}^R_{p\ell mn}\left(A_{\ell},A_{m,\text{noise}}A_n^{\ast}+A_mA_{n,\text{noise}}^{\ast}\right), \label{eq:NR}
\end{align} \label{eq:operators}
\end{subequations}
\endgroup 
where only the first-order noise terms are kept. $p$, $\ell$, $m$, and $n$ are eigenmode indices for the multimode field $\vec{\mathbb{A}}(\vec{r},t)=\sum_p\vec{F}_p(\vec{r}_{\perp})A_p(z,t)$, where $\vec{F}_j(\vec{r}_{\perp})$ is the normalized fiber eigenfield (in \si{1/\m}). Terms with and without $A_{p,\text{noise}}$ represent the spontaneous and the stimulated terms, respectively [Eq.~(\ref{eq:operators})]. Compared to the traditional shot-noise approach, the dispersion term of the noise, $\hat{\mathcal{D}}A_{p,\text{noise}}$, is removed from Eq.~(\ref{eq:partial_Ap}) to prevent artificial FWM resulting from unrealistically-consistent phase increment of the noise. Note that noise is still involved in FWM processes [Eq.~(\ref{eq:NK})]. $\hat{\mathcal{G}}_e$ is the stimulated amplification that considers only the emission cross sections. The Kerr and Raman terms in Eq.~(\ref{eq:operators}) follow
\begin{subequations}
\begin{align}
\hat{\mathcal{C}}^K_{p\ell mn}\left(A_{\ell}A_mA_n^{\ast}\right) & \sim\sum_{\ell mn}S^K_{p\ell mn}A_{\ell}A_mA_n^{\ast} \allowdisplaybreaks \label{eq:NKAp} \allowdisplaybreaks \\
\hat{\mathcal{C}}^R_{p\ell mn}\left(A_{\ell},A_mA_n^{\ast}\right) & \sim\sum_{\ell mn}S^R_{p\ell mn}A_{\ell}\left[R\ast\left(A_mA_n^{\ast}\right)\right] \label{eq:NRAp}
\end{align}
\end{subequations}
where $S^K_{p\ell mn}$ and $S^R_{p\ell mn}$ are overlap integrals of eigenfields \cite{Horak2012}, $R(t)$ is the Raman delayed response. There are three spontaneous Raman terms. $\hat{\mathcal{C}}^R_{p\ell mn}\left(A_{\ell},A_{m,\text{noise}}A_n^{\ast}+A_mA_{n,\text{noise}}^{\ast}\right)$ represents Raman frequency shifting for the signal field $A_{\ell}$, directly corresponding to spontaneous Raman generation at Stokes and anti-Stokes frequencies. On the other hand, $\hat{\mathcal{C}}^R_{p\ell mn}\left(A_{\ell,\text{noise}},A_mA_n^{\ast}\right)$ represents nonlinear-phase modulations and Raman frequency shifting for the weak noise field $A_{\ell,\text{noise}}$, so this term can be reasonably ignored. Since the spontaneous Raman term contains noise-involved nonlinear effects in addition to the Raman growth, such as Raman-induced phase modulations, the use of $\mathfrak{F}[R]$ correctly takes them into account. However, this induces loss at anti-Stokes frequencies, which should be gain as in the Langevin approach with $g^{AS}\propto\abs{\Im\left[\mathfrak{F}[R]\right]}\geq0$. Because spontaneous anti-Stokes generation is dominated by spontaneous FWM generation, involving pump and Stokes fields, which is correctly considered with $\mathfrak{F}[R]$, this does not affect the overall behavior. In fact, the modified shot-noise approach still creates spontaneous anti-Stokes fields in lossy environments, better simulating the reality than the shot-noise approach that induces constant absorption [Fig.~\ref{fig:models_loss}(b)]. Eq.~(\ref{eq:NRAp}) raises a question of whether $\left[n_{\text{th}}(\abs{\Omega})+\Theta(-\Omega)\right]$ should be added to the $\left[\mathfrak{F}[R]\mathfrak{F}[A_mA_n^{\ast}]\right]$, after applying the convolution theorem to Eq.~(\ref{eq:NRAp}), to consider the Stokes preference and thermal phonons in spontaneous Raman generation. This additional term does not correct the anti-Stokes loss into gain but reduces its loss value. Moreover, it changes the spontaneous FWM effect that dominates the spontaneous anti-Stokes growth. Since $n_{\text{th}}(\abs{\Omega})$ is generally small at Raman frequencies, omission of this term does not affect Stokes generation as well. More details are discussed in \ref{sec:Various noise models}.

\begin{figure}[!ht] 
\centering
\includegraphics[width=.95\linewidth]{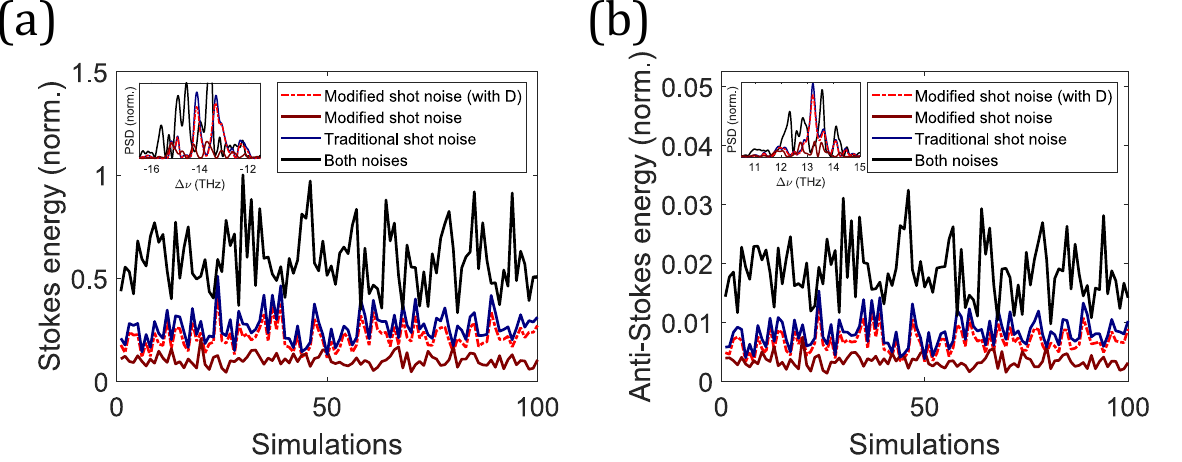}
\caption{Noise-model comparisons for spontaneous Raman (a) Stokes and (b) anti-Stokes generation with \num{100} simulations each. Spectra are shown in insets. PSD: power spectral density. The \SIadj{100}{\nano\joule} and \SIadj{10}{\ps}-long pulse at \SI{1030}{\nm} is launched into a \SIadj{1}{\m}-long single-mode silica fiber. Normalization of energies maintains the relative strength between two signals. Modified shot-noise approach closely follows the Raman Langevin approach. ``Modified shot noise (with $D$)'' includes the $\hat{\mathcal{D}}A_{p,\text{noise}}$ term in the UPPE, which creates the same outcome as the shot-noise approach (where their small difference comes from numerical implementation details) with boosted spontaneous generation arising from artificial noise phase matching.}
\label{fig:UPPE_noise}
\end{figure}

Unlike Raman scattering and amplification in rare-earth-doped fibers, shot noise is compatible with the noise-seeded Kerr effects [Eq.~(\ref{eq:NK})] due to its lack of aforementioned asymmetries. Thus, it has been customary to simultaneously incorporate shot noise (in the input field) and Raman Langevin term (in the governing equation) in modeling noise effects in both Kerr and Raman phenomena \cite{Dudley2006}. However, this approach doubles the vacuum fluctuations and the resulting generated Stokes photons (if unsaturated) [Fig.~\ref{fig:UPPE_noise}(a)]. Addition of shot noise elevates the noise floor at anti-Stokes frequencies, rendering spontaneous anti-Stokes generation in the Raman Langevin term redundant and amplifying anti-Stokes effects [Fig.~\ref{fig:UPPE_noise}(b)]. The consistent FWM effect of shot noise can also amplify anti-Stokes fields. These, combined, result in an artificially-enhanced Raman response to the examined physical phenomena.

To model noise-seeded processes, traditionally two approaches are employed. The \emph{traditional shot-noise approach} includes the noise photon to the input field, whose governing equation, however, consists of only stimulated terms. The \emph{stochastic approach} includes statistical Langevin terms to the governing equation, resulting in a stochastic differential equation \cite{Oeksendal2010}. Shot noise is easy to apply but it ignores the required asymmetries and exhibits an elevated noise floor. Stochastic approach is more accurate as it constantly considers decorrelation along $z$, but it is difficult to identify all spontaneous terms that are embedded in the stimulated governing equation as Langevin terms, such as numerous cascaded spontaneous FWM and Raman generation. Only the modified shot-noise approach can simulate the noise easily and consider the asymmetries in spontaneous generation, as well as free from an artificially-elevated noise floor. Readers are referred to \ref{sec:Various noise models} for illustrations of these three models and their comparison.

In conclusion, we present a new \emph{modified shot-noise} model that directly incorporates noise into the governing pulse propagation equation. It can accurately and easily solve for noise-seeded processes, even for Raman scattering and amplification in rare-earth-doped waveguides that exhibit asymmetries.

\begin{backmatter}
\bmsection{Funding} Office of Naval Research (N00014-19-1-2592); National Institutes of Health (R01EB033179, U01NS128660).

\bmsection{Disclosures} The authors declare no conflicts of interest.

\bmsection{Data availability} The code based on this model has been made publicly available at \url{https://github.com/AaHaHaa/MMTools} for solid-core fibers and \url{https://github.com/AaHaHaa/gas_UPPE} for hollow-core fibers.

\bmsection{Author comments} We currently plan to submit this paper only in preprint to ensure open access for readers. It is a small paper with a simple core concept of Eq.~(\ref{eq:partial_Ap}), which we do not think worth a hassle of journal peer review. Nonetheless, it presents significant foundational insights into noise-seeded processes. We developed this model while implementing the spontaneous Langevin terms to correctly consider spontaneous Raman scattering but found it numerically complicated and confusing. We couldn't find any existing documents discussing the relationship or numerical equivalence between shot noise and Langevin term, which leads to this paper and \ref{sec:Various noise models}. Hope that readers can enjoy and learn from it. Feel free to contact me via email with any comments or critiques.

\end{backmatter}

\bibliography{noise_modeling.bib}


\newpage
\onecolumn
\setcounter{equation}{0}
\renewcommand{\theequation}{A\arabic{equation}}
\setcounter{figure}{0}
\renewcommand{\thefigure}{A\arabic{figure}}
\setcounter{section}{0}
\renewcommand{\thesection}{Appendix \arabic{section}}
\section{Multimode unidirectional pulse propagation equation}
\label{sec:MM-UPPE}
Multimode unidirectional pulse propagation equation (MM-UPPE) with both the stimulated and spontaneous contributions (please find derivation in the supplement of \cite{Chen2024}):
\begin{align}
\partial_zA_p(z,\Omega) & =i\left[\beta_p(\omega)-\left(\beta_{(0)}+\beta_{(1)}\Omega\right)\right]A_p(z,\Omega) && \rightarrow\text{Dispersion} \nonumber \\
&\hspace{1em} +\hat{\mathcal{G}}A_p(z,\Omega)+\hat{\mathcal{G}}_eA_{p,\text{noise}}(z,\Omega) && \rightarrow\text{Gain} \nonumber \\
& \hspace{1em}+i\omega\kappa\sum_{\ell mn}\Bigg\{\kappa_eS^K_{p\ell mn}\mathfrak{F}\left[A_{\ell}A_mA_n^{\ast}+\left(A_{\ell,\text{noise}}A_mA_n^{\ast}+A_{\ell}A_{m,\text{noise}}A_n^{\ast}+A_{\ell}A_mA_{n,\text{noise}}^{\ast}\right)\right] && \rightarrow\text{Kerr} \nonumber \\
& \hspace{6.3em}
+S^{R_a}_{p\ell mn}\mathfrak{F}\left[A_{\ell}\left\{R_a\ast\left[A_mA_n^{\ast}+\left(A_{m,\text{noise}}A_n^{\ast}+A_mA_{n,\text{noise}}^{\ast}\right)\right]\right\}\right] && \rightarrow\text{isotropic Raman} \nonumber \\
& \hspace{6.3em}+S^{R_b}_{p\ell mn}\mathfrak{F}\left[A_{\ell}\left\{R_b\ast\left[A_mA_n^{\ast}+\left(A_{m,\text{noise}}A_n^{\ast}+A_mA_{n,\text{noise}}^{\ast}\right)\right]\right\}\right]\Bigg\} && \rightarrow\text{anisotropic Raman}, \label{eq:UPPE}
\end{align}
where the Raman term can be computed with the convolution theorem: $R_r(t)\ast X(t)=\frac{1}{C_{\mathfrak{F}}}\mathfrak{F}^{-1}\left[\mathfrak{F}\left[R_r\right]\mathfrak{F}\left[X(t)\right]\right]$. $A_p(z,T)$ is the electric field envelope (in \si{\sqrt{W}}) of mode $p$, whose Fourier transform is $A_p(z,\Omega)=\mathfrak{F}[A_p(z,T)]$. $z$ is the propagation distance. The Fourier transform is applied with respect to angular frequency $\Omega=\omega-\omega_0$, where $\omega_0$ is the center angular frequency of the numerical frequency window. $\beta_p$ is the propagation constant of the mode $p$. $\beta_{(0)}$ and $\beta_{(1)}$ are to reduce the propagating global-phase increment, thus allowing a larger step size to facilitate simulations, in which $\beta_{(1)}$ is the inverse group velocity of the moving reference frame that introduces the delayed time $T=t-\beta_{(1)}z$. $\left(\cdot\right)_{\hat{\mathcal{G}}}$ represents that this derivative comes from the gain evolution \cite{Chen2023a,Chen2025}. $\kappa=1/\left(\epsilon_0^2\left[n_{\text{eff}}(\omega)\right]^2c^2\right)$ with $c$ the speed of light in vacuum. $\kappa_e=\left(3/4\right)\epsilon_0\chi^{(3)}_{\text{electronic}}(\omega)$ with $\chi^{(3)}_{\text{electronic}}(\omega)$ the third-order nonlinear susceptibility of the electronic response (in \si{\m^2/\V^2}). $R_a(t)$ and $R_b(t)$ are isotropic and anisotropic Raman response functions \cite{Stolen1989,Lin2006,Chen2024}. $S^K_{p\ell mn}$, $S^{R_a}_{p\ell mn}$, and $S^{R_b}_{p\ell mn}$ follow
\begingroup\allowdisplaybreaks
\begin{subequations}
\begin{align}
S^K_{p\ell mn} & =\frac{2}{3}S^{R_a}_{p\ell mn}+\frac{1}{3}S^k_{p\ell mn} \label{eq:SK} \\
S^{R_a}_{p\ell mn} & =\int\left(\vec{F}_p^{\ast}\cdot\vec{F}_{\ell}\right)\left(\vec{F}_m\cdot\vec{F}_n^{\ast}\right)\diff^2x \label{eq:SRa} \\
S^{R_b}_{p\ell mn} & =\frac{1}{2}\left[\int\left(\vec{F}_p^{\ast}\cdot\vec{F}_m\right)\left(\vec{F}_{\ell}\cdot\vec{F}_n^{\ast}\right)\diff^2x+S^k_{p\ell mn}\right], \label{eq:SRb}
\end{align}
\end{subequations}
\endgroup
where $S^k_{p\ell mn}=\int\left(\vec{F}_p^{\ast}\cdot\vec{F}_n^{\ast}\right)\left(\vec{F}_{\ell}\cdot\vec{F}_m\right)\diff^2x$. $\diff^2x=\diff x\diff y$ represents the integral over the spatial domain.

MM-UPPE in Eq.~(\ref{eq:UPPE}) is a general formulation, which is widely used in gas-filled hollow-core fiber. However, in solid-core fibers, Raman model with a Raman fraction $f_R$ representing the contribution of the Raman response of all $\chi^{(3)}$ nonlinearities is more-commonly used. This results from the short dephasing time of solid's Raman responses, always leading to Raman-enhanced self-phase modulation (SPM) \cite{Chen2024}. In this representation,
\begin{subequations}
\begin{align}
\kappa\kappa_e & =\kappa\left\{\frac{3}{4}\epsilon_0\left[\left(1-f_R\right)\chi^{(3)}_{xxxx}\right]\right\}=\kappa\left\{\frac{3}{4}\epsilon_0\left[\left(1-f_R\right)\frac{4\epsilon_0\left[n_{\text{eff}}(\omega)\right]^2c}{3}n_2\right]\right\}=\frac{n_2}{c}\left(1-f_R\right) \\
\kappa R_a(t) & =\kappa\left[\epsilon_0\left(f_Rf_a\chi^{(3)}_{xxxx}\right)\frac{3}{4}h_a(t)\right]=\kappa\left[f_Rf_a\epsilon_0^2\left[n_{\text{eff}}(\omega)\right]^2cn_2h_a(t)\right]=\frac{n_2}{c}f_Rf_ah_a(t) \\
\kappa R_b(t) & =\kappa\left[\epsilon_0\left(f_Rf_b\chi^{(3)}_{xxxx}\right)\frac{3}{4}h_b(t)\right]=\kappa\left[f_Rf_b\epsilon_0^2\left[n_{\text{eff}}(\omega)\right]^2cn_2h_b(t)\right]=\frac{n_2}{c}f_Rf_bh_b(t),
\end{align}
\end{subequations}
where $n_2$ is the ``total'' nonlinear refractive index, including electronic and Raman nonlinearities, which is measured usually by the amount of SPM a pulse gains after passing through the medium. $f_a$ and $f_b$ are Raman fractions of the total Raman response for isotropic and anisotropic Raman responses [$h_a(t)$ and $h_b(t)$], respectively ($f_a+f_b=1$) \cite{Stolen1989,Lin2006}. $h_a$ and $h_b$ are normalized such that $\int_0^{\infty} h_a(t)\diff t=\int_0^{\infty} h_b(t)\diff t=1$.

For details about the doped-ion amplification contributions, both stimulated $\hat{\mathcal{G}}A_p(z,\Omega)$ and spontaneous $\hat{\mathcal{G}}_eA_{p,\text{noise}}(z,\Omega)$, please see \cite{Chen2023a,Chen2025}.

\newpage
\section{Various noise models}
\label{sec:Various noise models}
\begin{figure}[!ht] 
\centering
\includegraphics[width=.8\linewidth]{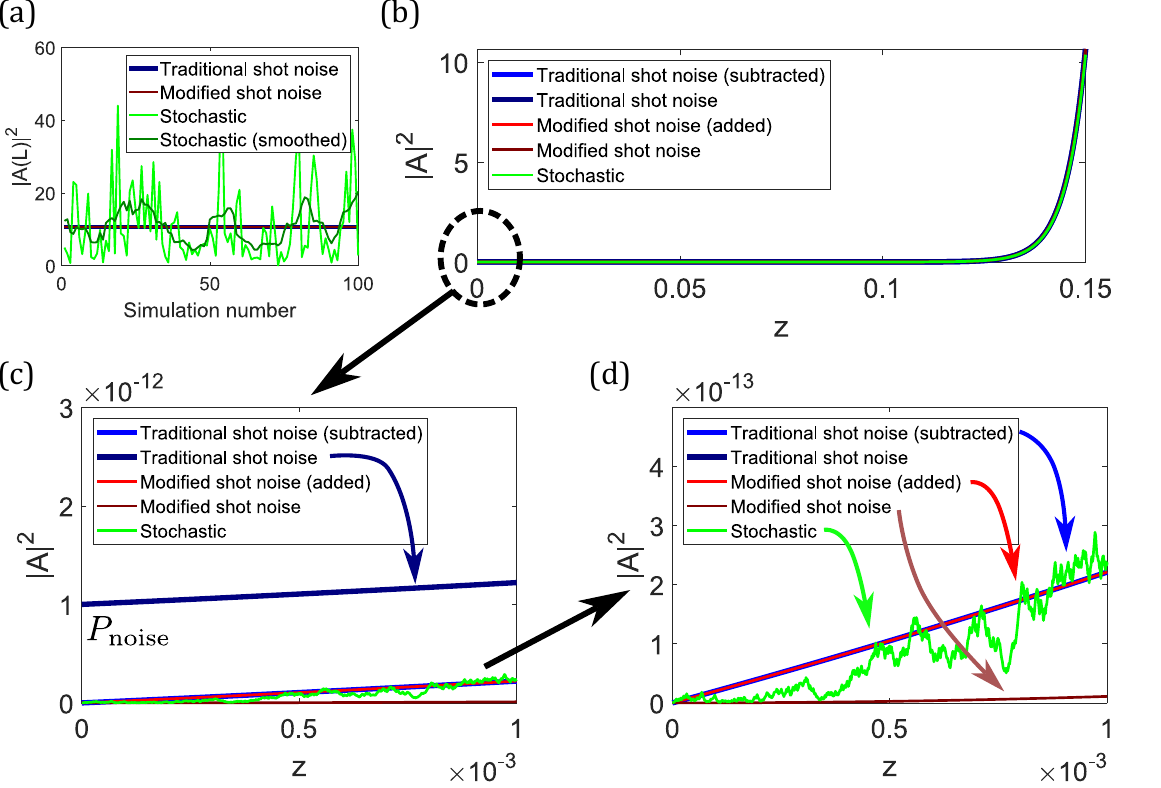}
\caption{Noise-model studies with $g=200$ and the vacuum-fluctuation noise floor $P_{\text{noise}}=10^{-12}$ [Eq.~(\ref{eq:dAdz})]. We let $\mathfrak{R}=\Exp^{i\phi}$ vary only over phases for the demonstration purpose of having consistent intensities over different simulations. (a) \num{100} simulations of the three models. The line from the stochastic model (light green) is smoothed (dark green) for a clearer comparison. (b) Power evolutions. In addition to the original powers obtained from each model ($P_j=\abs{A_j}^2$, where $j=s,s',\text{sto}$ for shot-noise [Eq.~(\ref{eq:shot_dAdz})], modified shot-noise [Eq.~(\ref{eq:modified_shot_dAdz})], and stochastic models [Eq.~(\ref{eq:stochastic_dAdz})], respectively), the power of ``the shot-noise model (subtracted)'' (light-blue line) follows $P_{\text{shot noise (subtracted)}}(z)=P_{s}(z)-P_{\text{noise}}$ and the power of ``the modified shot-noise model (added)'' (red line) follows Eq.~(\ref{eq:P_modified_shot_noise}). (c) and (d) are close-views of (b). We can see that there is always a background noise in the shot-noise approach (dark-blue line), and the power $P_{s'}$ directly from the modified shot-noise model (brown line) is not only weak but also quadratically-increasing at a small $z$.}
\label{fig:models_gain}
\end{figure}

In this section, we discuss some basic concepts of various noise models, including the validation of the one-photon-per-mode shot-noise model. Additionally, we address the question of whether it is feasible to eliminate its initially-added noise floor while retaining only the noises that arise from it, \ie those from the governing equation. The conclusion of the discussion leads to the proposed noise-including governing equation in the article. Fig.~\ref{fig:models_gain} concludes the discussion, which will be detailed in the following paragraphs.

We assume a simple governing equation:
\begin{equation}
\dod{A}{z}(z)=\frac{g}{2}A(z),\text{ with }A(0)=0,
\label{eq:dAdz}
\end{equation}
where $g\in\mathbb{R}$ represents the constant undepleted gain. To simplify the analysis, we assume that there is no initial field $A(0)=0$. A shot-noise model treats the stochastic effect with the noise-added input field $A_s(0)=A_{\text{noise}}$ (denoted with $s$):
\begin{equation}
\dod{A_{s}}{z}(z)=\frac{g}{2}A_{s}(z),\text{ with }A_{s}(0)=A_{\text{noise}}.
\label{eq:shot_dAdz}
\end{equation}
Consequently, the output field is given by $A_{s}(z)=\Exp^{gz/2}A_{\text{noise}}$ leading to a power defined by $P_{s}(z)=\abs{A_{s}(z)}^2=\Exp^{gz}\abs{A_{\text{noise}}}^2\approx\left(1+gz\right)P_{\text{noise}}$ with $P_{\text{noise}}=\abs{A_{\text{noise}}}^2$ a constant. The linear increase of the term $gzP_{\text{noise}}$ mimics the initial linear \emph{power} growth of SE [Fig.~\ref{fig:models_gain}(c)]. While it is possible to isolate the power contribution from spontaneous generation, represented as $P(z)=gzP_{\text{noise}}$ during the numerical post-processing stage following a simulation, it may be tempting to remove it at an earlier stage. This can be achieved by having a modified governing equation  (denoted with $s'$) such that the noise effect manifests itself only during the evolution:
\begin{equation}
\dod{A_{s'}}{z}(z)=\frac{g}{2}\left(A_{s'}(z)+A_{\text{noise}}\right),\text{ with }A_{s'}(0)=0.
\label{eq:modified_shot_dAdz}
\end{equation}
At a small $z$, the stimulated effect is weak and thus the field follows $A_{s'}(z)=\sum_{n=1}^M\frac{g\triangle z}{2}A_{\text{noise}}$ with $z=M\triangle z$. This suggests that $A_{\text{noise}}$ needs to be a constant, rather than a random variable that varies along $z$; otherwise, the averaged power $\expval{P_{s'}(z)}=\sum_{n=1}^M\expval{\abs{\frac{g\triangle z}{2}A_{\text{noise}}}^2}=M\left(\frac{g\triangle z}{2}\right)^2P_{\text{noise}}=\frac{1}{M}\left(\frac{gz}{2}\right)^2P_{\text{noise}}$ will depend on $M$, and thus the numerical step size $\triangle z$. Since $A_{\text{noise}}$ is constant, $A_{s'}(z)$ from the modified shot-noise approach [Eq.~(\ref{eq:modified_shot_dAdz})] is the same as the noise-removed $A_{s}(z)-A_{\text{noise}}$ from the shot-noise approach [Eq.~(\ref{eq:shot_dAdz})]. Therefore, it does not need post-processing noise removal. This is the approach we proposed earlier in the article.

It is worth noting the mathematical requirement to calculate the power when $\abs{A}^2\ll P_{\text{noise}}$. The power evolution from the shot-noise approach follows
\begin{align}
P_{s}(z) & =\abs{A_{s'}(z)+A_{\text{noise}}}^2 \nonumber \\
& = \abs{A_{s'}(z)}^2+P_{\text{noise}}+2\Re[A^{\ast}_{s'}(z)A_{\text{noise}}],\quad\text{note that }A_{s'}(z)\approx\frac{gz}{2}A_{\text{noise}}\text{ as }\abs{z}\ll1 \nonumber \\
& \approx\abs{\frac{gz}{2}A_{\text{noise}}}^2+P_{\text{noise}}+gz\abs{A_{\text{noise}}}^2=\left(\frac{g^2}{4}z^2+1+gz\right)P_{\text{noise}}.
\end{align}
At a small $z$, the $z^2$ term can be discarded, leading to the linear growth of spontaneous generation in the shot-noise approach, as expected from physics. However, in the modified shot-noise approach [Eq.~(\ref{eq:modified_shot_dAdz})], its power, at a small $z$, follows $P_{s'}(z)=\abs{A_{s'}(z)}^2\propto z^2$, leading to a small quadratically-increasing power [Fig.~\ref{fig:models_gain}(d)]. Therefore, to compute the power correctly in the modified shot-noise approach, especially at a low power, it is necessary to include the noise before absolute square, followed by its removal:
\begin{equation}
P_{s'}(z)=\abs{A_{s'}(z)+A_{\text{noise}}}^2-P_{\text{noise}}.
\label{eq:P_modified_shot_noise}
\end{equation}

The noise effect of Eq.~(\ref{eq:dAdz}) can be modeled most correctly with the stochastic governing equation with a Langevin term, due to its constant stochastic contributions over $z$:
\begin{equation}
\dod{A_{\text{sto}}}{z}(z)=\frac{g}{2}A_{\text{sto}}(z)+\dod{A_{\text{spon}}}{z}(z),\text{ with }A_{\text{sto}}(0)=0,
\label{eq:stochastic_dAdz}
\end{equation}
whose Langevin term $\od{A_{\text{spon}}}{z}$ follows the Dirac-delta-correlated function $\expval{\dod{A_{\text{spon}}}{z}(z)\left(\dod{A_{\text{spon}}}{z}(z')\right)^{\ast}}=gP_{\text{noise}}\delta(z-z')$ and has a zero mean, $\expval{\dod{A_{\text{spon}}}{z}(z)}=0$ (see \ref{sec:Langevin}). In fact, this is the correct mathematical formulation of the governing equation to display the existence of a noise effect during evolution, whereas Eq.~(\ref{eq:dAdz}) consists of only the stimulated term. We can derive the spontaneous increment from the following (or, equivalently, from the correlation function of the Langevin term):
\begin{align}
\dod{\abs{A}^2}{z}(z)=g\abs{A(z)}^2\qquad & \Rightarrow\qquad\dod{\expval{\abs{A_{\text{spon}}}^2}}{z}(z)=gP_{\text{noise}} \nonumber \\
&\Rightarrow\qquad\diff A_{\text{spon}}(z)=\sqrt{gP_{\text{noise}}\diff z}\mathfrak{R}(z),
\end{align}
with $\mathfrak{R}(z)$ the complex random number following the joint normal distribution. During numerical computations, $A_{\text{sto}}(z+\diff z)=A_{\text{sto}}(z)+\diff A_{\text{sto}}$ where
\begin{equation}
\diff A_{\text{sto}}=\frac{g}{2}A_{\text{sto}}(z)\diff z+\diff A_{\text{spon}}(z)=\frac{g}{2}A_{\text{sto}}(z)\diff z+\sqrt{gP_{\text{noise}}\diff z}\mathfrak{R}(z).
\end{equation}
Since the stochastic model considers the decorrelation along $z$ [due to $\mathfrak{R}(z)$], it exhibits a downside, compared to both the aforementioned shot-noise approaches, of necessitating a small numerical increment $\triangle z$. Otherwise, the spontaneous generation will not follow the initial linear power growth arising from its randomness.

Figs.~\ref{fig:models_gain}(a) and (b) show the consistency between three models we discussed: the traditional shot-noise model [Eq.~(\ref{eq:shot_dAdz})], the modified shot-noise model [Eq.~(\ref{eq:modified_shot_dAdz})], and the stochastic model [Eq.~(\ref{eq:stochastic_dAdz})]. They give the same output at high powers when the background-noise subtraction or addition does not play a crucial role [Eq.~(\ref{eq:P_modified_shot_noise})]. Here, we demonstrate noise models with the simplest single-noise term. Despite the accuracy of the stochastic model, in general, it is difficult to explicitly express all noise terms stochastically in the governing equation, which include, for example, spontaneous FWM generation and Stokes and anti-Stokes generation and their unlimited cascaded processes. Therefore, in practice, it is desirable to apply the noise elsewhere that can interact with the established stimulated effects, which leads to ``shot noise.'' Two shot-noise approaches include all noise effects with a simple inclusion of one-noise-photon CW background. Compared to the traditional shot-noise model, the modified shot-noise approach is preferred, as it can deal with the asymmetries exhibiting in the spontaneous generation as described in the article.

\begin{figure}[!ht] 
\centering
\includegraphics[width=.8\linewidth]{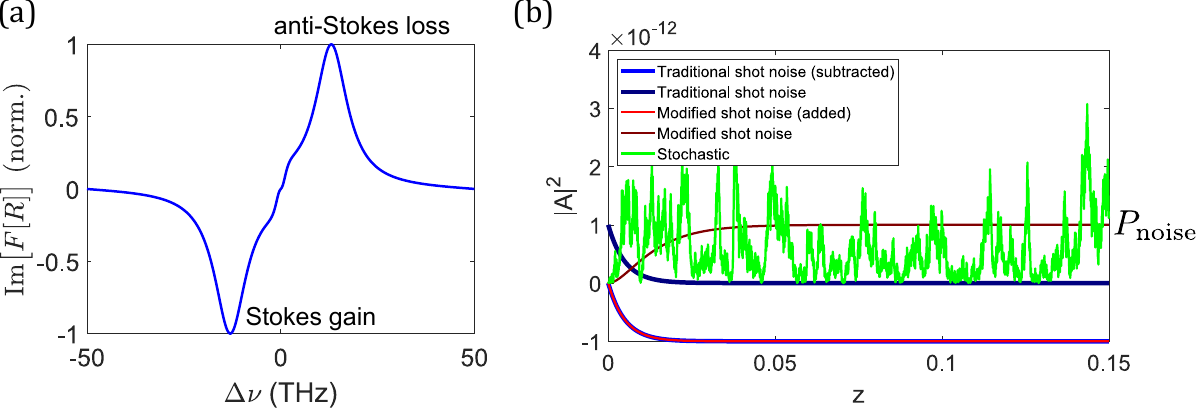}
\caption{(a) Raman gain/loss ($g_R\propto-\Im\left[\mathfrak{F}[R]\right]$; the negative sign results from the extra ``$i$'' in the nonlinear prefactor [Eq.~(\ref{eq:UPPE})]) in silica. (b) Noise-model studies with $g=-200$ and other conditions the same as in Fig.~\ref{fig:models_gain}. }
\label{fig:models_loss}
\end{figure}

In the article, we demonstrate that $\mathfrak{F}[R]$ results in loss rather than gain at anti-Stokes frequencies [Fig.~\ref{fig:models_loss}(a)]. In the Langevin approach, the anti-Stokes gain $g^{AS}$ is accurately represented as $g^{AS}\propto\abs{\Im\left[\mathfrak{F}[R]\right]}\geq0$ with an absolute square. Consequently, this raises the question of whether $\mathfrak{F}[R]$ should be substituted with $-i\abs{\Im\left[\mathfrak{F}[R]\right]}$ to ensure positive Raman gain at all frequencies. This substitution, however, removes all the other nonlinear effects except for the Raman gain. Unlike the traditional shot-noise approach that constantly induces stimulated anti-Stokes loss, our modified shot-noise approach generates a field that saturates at $P_{\text{noise}}$ [Fig.~\ref{fig:models_loss}(b)] due to $A_s=A_{s'}+A_{\text{noise}}$. This modified shot-noise framework permits spontaneous generation even in absorptive environments, thereby providing a more realistic simulation and outperforming the traditional shot-noise approach. Despite the unrealistic saturation, the generated field serves as a seed for subsequent stimulated amplification, effectively producing more anti-Stokes fields. Furthermore, since spontaneous anti-Stokes generation is predominantly driven by the FWM process involving pump and Stokes fields, it is essential to accurately simulate FWM with $\mathfrak{F}[R]$, rather than incorrectly prioritizing weak gain associated with $\Im\left[\mathfrak{F}[R]\right]$ while neglecting other critical effects. As a result, despite seemingly anti-Stokes loss, the modified shot-noise approach can accurately model spontaneous anti-Stokes generation.

Because spontaneous anti-Stokes generation is dominated by the spontaneous FWM generation, traditional shot noise can still effectively create anti-Stokes fields despite its constant loss from the $\Im\left[\mathfrak{F}[R](\omega_R)\right]>0$ contribution ($\omega_R$: Raman transition frequency).

\newpage
\section{Fourier transform}
\label{sec:Fourier_Transform}
Here, we briefly describe the Fourier transform (FT) and discrete Fourier transform (DFT), and how to compute physical quantities with physical-useful units. For a complete discussion, please refer to \cite{Chen2025a}.

\subsection{Definition}
\label{subsec:Definition}
FT can be defined differently whose constants might differ in different conventions, such as for the convolution theorem:
\begin{equation}
\begin{aligned}
\mathfrak{F}[A*B] & =\frac{1}{C_{\mathfrak{F}}}\mathfrak{F}[A]\mathfrak{F}[B] \\
\mathfrak{F}^{-1}[A*B] & =\frac{1}{C_{\mathfrak{IF}}}\mathfrak{F}^{-1}[A]\mathfrak{F}^{-1}[B]
\end{aligned}
\quad\text{ if }\quad
\begin{aligned}
A(\omega)=\mathfrak{F}[A](\omega) & = C_{\mathfrak{F}}\int^{\infty}_{-\infty}A(t)\Exp^{i\omega t}\diff t\\
A(t)=\mathfrak{F}^{-1}[A](t) & = C_{\mathfrak{IF}}\int^{\infty}_{-\infty}A(\omega)\Exp^{-i\omega t}\diff\omega
\end{aligned},
\label{eq:Fourier_Transform}
\end{equation}
whose discrete counterpart is
\begin{equation}
\begin{aligned}
\mathfrak{F}_{D_c}[A*B] & =\frac{1}{C_{\mathfrak{F}}}\mathfrak{F}_{D_c}[A]\mathfrak{F}_{D_c}[B] \\
\mathfrak{F}_{D_c}^{-1}[A*B] & =\frac{1}{C_{\mathfrak{IF}}}\mathfrak{F}_{D_c}^{-1}[A]\mathfrak{F}_{D_c}^{-1}[B]
\end{aligned}
\quad\text{ if }\quad
\begin{aligned}
A(\omega)=\mathfrak{F}_{D_c}[A](\omega) & = C_{\mathfrak{F}}\sum_{n=1}^{\mathfrak{N}}A(t_n)\Exp^{i\omega t_n}\triangle t\\
A(t)=\mathfrak{F}_{D_c}^{-1}[A](t) & = C_{\mathfrak{IF}}\sum_{n=1}^{\mathfrak{N}}A(\omega_n)\Exp^{-i\omega_nt}\triangle\omega
\end{aligned},
\label{eq:discrete_Fourier_Transform_c}
\end{equation}
where $C_{\mathfrak{F}}C_{\mathfrak{IF}}=\frac{1}{2\pi}$. $\mathfrak{F}$ and $\mathfrak{F}_{D_c}$ are the continuous and discrete versions of Fourier transform, respectively. $\triangle w=2\pi\triangle\nu=\frac{2\pi}{\mathfrak{N}\triangle t}$. In the discrete manner, $t_n=n\triangle t$ and $\omega_n=n\triangle\omega$. If the sampling frequency is high enough, $\mathfrak{F}_{D_c}[\cdot]\approx\mathfrak{F}[\cdot]$ and $\mathfrak{F}_{D_c}^{-1}[\cdot]\approx\mathfrak{F}^{-1}[\cdot]$.

In practice, during numerical computations, we treat the result from the following DFT simply as FT, which differs from Eq.~(\ref{eq:discrete_Fourier_Transform_c}) in constants and units:
\begin{equation}
\begin{aligned}
A_D(\omega)=\mathfrak{F}_D[A](\omega) & = C_{\mathfrak{F}_D}\sum_{n=1}^{\mathfrak{N}}A(t_n)\Exp^{i\omega t_n}\\
A(t)=\mathfrak{F}_D^{-1}[A_D](t) & = C_{\mathfrak{IF}_D}\sum_{n=1}^{\mathfrak{N}}A_D(\omega_n)\Exp^{-i\omega_nt},
\label{eq:discrete_Fourier_Transform}
\end{aligned}
\end{equation}
where $C_{\mathfrak{F}_D}C_{\mathfrak{IF}_D}=\frac{1}{\mathfrak{N}}$ and $\mathfrak{N}$ is the number of discrete points. The Fourier-Transform $A_D(\omega_n)$ is denoted with an extra ``$D$'' subscript to distinguish it from $A(\omega_n)$ of Eqs.~(\ref{eq:Fourier_Transform}) and (\ref{eq:discrete_Fourier_Transform_c}). Therefore, it is important to derive a relationship between $A(\omega_n)$ and $A_D(\omega_n)$, which follows
\begin{equation}
\frac{1}{C_{\mathfrak{F}_D}}A_D(\omega)\triangle t=\frac{1}{C_{\mathfrak{F}}}A(\omega).
\label{eq:AD-A_transform}
\end{equation}
For the commonly-used Fourier-Transform convention in the laser field (and the one we use in this article) such that $C_{\mathfrak{F}_D}=\frac{1}{\mathfrak{N}}$, Eq.~(\ref{eq:AD-A_transform}) becomes $A_D(\omega)=A(\omega)\triangle\nu/C_{\mathfrak{F}}$.

\subsection{Relations between FT and DFT}
\label{subsec:Relations_between_FT_and_DFT}
In this section, we derive several formulae for conversion of physical quantities between Fourier transform and discrete Fourier transform.
\begin{align}
\int_{-\infty}^{\infty}\abs{A(t)}^2\diff t & =\int_{-\infty}^{\infty}\left(C_{\mathfrak{IF}}\right)^2\int_{-\infty}^{\infty}\int_{-\infty}^{\infty}A(\omega)A^{\ast}(\omega')\Exp^{-i\left(\omega-\omega'\right)t}\diff\omega\diff\omega'\diff t \nonumber \\
& =\left(C_{\mathfrak{IF}}\right)^2\int_{-\infty}^{\infty}\int_{-\infty}^{\infty}A(\omega)A^{\ast}(\omega')\left[\int_{-\infty}^{\infty}\Exp^{-i\left(\omega-\omega'\right)t}\diff t\right]\diff\omega\diff\omega' \nonumber \\
& =\left(C_{\mathfrak{IF}}\right)^2\int_{-\infty}^{\infty}\int_{-\infty}^{\infty}A(\omega)A^{\ast}(\omega')\left[2\pi\delta(\omega-\omega')\right]\diff\omega\diff\omega' \nonumber \\
& =2\pi\left(C_{\mathfrak{IF}}\right)^2\int_{-\infty}^{\infty}\abs{A(\omega)}^2\diff\omega \nonumber \\
& =\frac{C_{\mathfrak{IF}}}{C_{\mathfrak{F}}}\int_{-\infty}^{\infty}\abs{A(\omega)}^2\diff\omega\quad\because C_{\mathfrak{IF}}=\frac{1}{2\pi C_{\mathfrak{F}}} \label{eq:Parseval_derivation}
\end{align}
Eq.~(\ref{eq:Parseval_derivation}) leads to the general formulation of the Parseval's theorem:
\begin{equation}
\frac{1}{C_{\mathfrak{IF}}}\int_{-\infty}^{\infty}\abs{A(t)}^2\diff t=\frac{1}{C_{\mathfrak{F}}}\int_{-\infty}^{\infty}\abs{A(\omega)}^2\diff\omega
\label{eq:Parseval}
\end{equation}

Since $\abs{A(t)}^2$ has the unit of ``\si{\W}=\si{\joule/\s}'', $\frac{C_{\mathfrak{IF}}}{C_{\mathfrak{F}}}\abs{A(\omega)}^2=\frac{1}{2\pi C_{\mathfrak{F}}^2}\abs{A(\omega)}^2$ has the unit of ``\si{\joule/(\radian\cdot\Hz)}'' ($\omega$ has a unit of ``$\si{\Hz}/(2\pi)=\si{\radian\cdot\Hz}$''). To calculate the spectrum with the unit of ``\si{\joule/\Hz}'' numerically,
\begin{align}
P(\nu)=2\pi P(\omega)=\frac{1}{C_{\mathfrak{F}}^2}\abs{A(\omega)}^2=\left(\frac{\triangle t}{C_{\mathfrak{F}_D}}\right)^2\abs{A_D(\omega)}^2,
\label{eq:Pnu}
\end{align}
by applying Eq.~(\ref{eq:AD-A_transform}) and $\int P(\nu)\diff\nu=\int P(\omega)\diff\omega$. With the Fourier-Transform convention we use here ($C_{\mathfrak{F}_D}=\frac{1}{\mathfrak{N}}$), it becomes $P(\nu)=\abs{A_D(\omega)}^2\left(\mathfrak{N}\triangle t\right)^2=\abs{A_D(\omega)}^2\left(T^w\right)^2$, where $T^w$ is the time window.

In the common model of adding noise photon (\eg shot noise), the noise is added as one noise photon per frequency mode/bin (\si{\watt/\Hz}) $P_{\text{noise}}(\nu)=h\nu$ whose energy in a time window $T^w=\frac{1}{\triangle\nu}$ is $h\nu T^w=\frac{1}{C_{\mathfrak{F}}^2}\abs{A_{\text{noise}}(\omega)}^2$ \cite{Eggleston1980,Dudley2006,Frosz2010,Genier2019}, so 
\begin{equation}
\abs{A_{\text{noise}}(\omega)}=C_{\mathfrak{F}}\sqrt{T^wh\nu}.
\label{eq:noise}
\end{equation}
The discrete counterpart $A_{D,\text{noise}}(\omega)$, by the use of Eq.~(\ref{eq:AD-A_transform}), is
\begin{equation}
\abs{A_{D,\text{noise}}(\omega)}=\frac{C_{\mathfrak{F}_D}}{\triangle t}\sqrt{T^wh\nu},
\label{eq:noise_D}
\end{equation}
which leads to $\sqrt{h\nu\triangle\nu}$ (because $\triangle\nu=1/T^w$) with the Fourier-Transform convention we use here.

\newpage
\section{Langevin terms}
\label{sec:Langevin}
In this section, we would like to show the Langevin formulation of several noise processes.

\subsection{Equal-frequency/time correlation}
\label{subsubsec:Equal-frequency_correlation}
If a stochastic governing equation shows a field evolution independent of itself, it follows the Langevin equation:
\begin{equation}
\dod{A}{z}(z,\omega)=a\eta(z,\omega), \label{eq:dAdz_aeta}
\end{equation}
whose $a\in\mathbb{R}$ and $\eta(z,\omega)$ satisfies
\begin{subequations}
\begin{align}
\expval{\eta(z,\omega)} & =0 \\
\expval{\eta(z,\omega)\eta^{\ast}(z',\omega)} & =C(\omega)\delta(z-z'). \label{eq:etaetastar}
\end{align}
\end{subequations}
Note that frequency $\omega$ here can be substituted with time $t$. With $A(z,\omega)=a\int_0^z\eta(\xi,\omega)\diff\xi$ for $A(0,\omega)=0$, we obtain
\begin{align}
\expval{\abs{A(z,\omega)}^2} & =\expval{a^2\left(\int_0^z\eta(\xi,\omega)\diff\xi\right)\left(\int_0^z\eta^{\ast}(\xi',\omega)\diff\xi'\right)} \nonumber \\
& =a^2\int_0^z\int_0^z\expval{\eta(\xi,\omega)\eta^{\ast}(\xi',\omega)}\diff\xi\diff\xi' \nonumber \\
& =a^2C(\omega)\int_0^z\int_0^z\delta(\xi-\xi')\diff\xi\diff\xi' \nonumber \\
& =a^2C(\omega)z,
\label{eq:expval_A2}
\end{align}
which corresponds to the linear (spontaneous) growth of $\expval{\abs{A}^2}$ over $z$:
\begin{equation}
\dod{\expval{\abs{A(z,\omega)}^2}}{z}=a^2C(\omega).
\end{equation}
This shows that $A(z,\omega)$ follows the complex-number joint normal (Gaussian) distribution with zero mean and $a^2Cz$ variance, whose amplitude follows normal distribution
\begin{equation}
\mathcal{N}(\abs{A(z,\omega)}=\abs{A}\big|A(0,\omega)=0)=\frac{1}{\sqrt{2\pi\left(a^2C(\omega)z\right)}}\Exp^{-\frac{\abs{A}^2}{2\left(a^2C(\omega)z\right)}}
\end{equation}
and its phase is uniformly distributed within \numrange{0}{2\pi}.

\subsection{Correlation of interacting fields}
\label{subsubsec:Correlation_of_interacting_fields}
Some noise-seeded nonlinear processes occur due to interactions with the existing fields, in contrast to Eq.~(\ref{eq:dAdz_aeta}). It typically follows
\begin{equation}
\dod{A}{z}(z,t)=a\Gamma(z,t)A(z,t),
\label{eq:dAdz_Gamma}
\end{equation}
where $a\in\mathbb{R}$ and $\Gamma(z,t)$ satisfies
\begin{equation}
\expval{\Gamma(z,\omega)\Gamma^{\ast}(z',\omega')}=C(\omega)\delta(z-z')\delta(\omega-\omega').
\label{eq:GammaGamma}
\end{equation}
Eq.~(\ref{eq:dAdz_Gamma}) can be written in the frequency domain, leading to
\begin{align}
\dod{A}{z}(z,\omega) & =a\mathfrak{F}\left[\Gamma(z,t)A(z,t)\right] \nonumber \\
& =aC_{\mathfrak{IF}}\Gamma(z,\omega)\ast A(z,\omega)
\end{align}
by use of the convolution theorem. Next we follow the similar process of Eq.~(\ref{eq:expval_A2}) and obtain
\begin{align}
& \expval{\abs{A(z,\omega)}^2} \nonumber \\
& =\expval{\left(\int_0^zaC_{\mathfrak{IF}}\Gamma(\xi,\omega)\ast A(\xi,\omega)\diff\xi\right)\left(\int_0^zaC_{\mathfrak{IF}}\Gamma^{\ast}(\xi',\omega)\ast A^{\ast}(\xi',\omega)\diff\xi'\right)} \nonumber \\
& =\expval{a^2C_{\mathfrak{IF}}^2\int_0^z\int_0^z\left(\int_{-\infty}^{\infty}\Gamma(\xi,\omega')A(\xi,\omega-\omega')\diff\omega'\int_{-\infty}^{\infty}\Gamma^{\ast}(\xi',\omega'')A^{\ast}(\xi',\omega-\omega'')\diff\omega''\right)\diff\xi\diff\xi'} \nonumber \\
& =a^2C_{\mathfrak{IF}}^2\int_0^z\int_0^z\int_{-\infty}^{\infty}\int_{-\infty}^{\infty}\expval{\Gamma(\xi,\omega')\Gamma^{\ast}(\xi',\omega'')}A(\xi,\omega-\omega')A^{\ast}(\xi',\omega-\omega'')\diff\omega'\diff\omega''\diff\xi\diff\xi'\nonumber \\
& =a^2C_{\mathfrak{IF}}^2\int_0^z\int_0^z\int_{-\infty}^{\infty}\int_{-\infty}^{\infty}C(\omega')\delta(\xi-\xi')\delta(\omega'-\omega'')A(\xi,\omega-\omega')A^{\ast}(\xi',\omega-\omega'')\diff\omega'\diff\omega''\diff\xi\diff\xi'\nonumber \\
& =a^2C_{\mathfrak{IF}}^2\int_0^z\int_{-\infty}^{\infty}C(\omega')\abs{A(\xi,\omega-\omega')}^2\diff\omega'\diff\xi\nonumber \\
& =a^2C_{\mathfrak{IF}}^2\int_0^zC(\omega)\ast\abs{A(\xi,\omega)}^2\diff\xi \nonumber \\
& \approx a^2C_{\mathfrak{IF}}^2\left(C(\omega)\ast\abs{A(\omega)}^2\right)z,\quad\text{assume no pump depletion},
\label{eq:expval_A2_Gamma}
\end{align}
which leads to
\begin{equation}
\dod{\expval{\abs{A(z,\omega)}^2}}{z}\approx C_{\mathfrak{IF}}^2\left(C(\omega)\ast\abs{A(\omega)}^2\right).
\label{eq:dA2dz_Gamma}
\end{equation}

A typical example for Eq.~(\ref{eq:dAdz_Gamma}) is the spontaneous Raman scattering. Assume that there is no intrapulse Raman scattering and $C(\omega)=C_R(\omega)\delta(\omega+\omega_R)$ represents a prominent Stokes gain peak at $\omega_R$. We obtain
\begin{equation}
\dod{\expval{\abs{A(z,\omega^S)}^2}}{z}=C_{\mathfrak{IF}}^2C_R(-\omega_R)\abs{A(\omega^P=\omega^S+\omega_R)}^2.
\end{equation}

It is noteworthy that Eq.~(\ref{eq:dA2dz_Gamma}) is satisfied also for the complex-conjugate modification of Eq.~(\ref{eq:dAdz_Gamma}):
\begin{equation}
\dod{A}{z}(z,t)=a\Gamma(z,t)A^{\ast}(z,t).
\label{eq:dAastdz_Gamma}
\end{equation}

\end{document}